\documentclass[preprint,showpacs,preprintnumbers]{revtex4}
\usepackage{graphicx,amsmath}

\def\EQ{\begin{eqnarray}}
\def\EN{\end{eqnarray}}
\def\tit{{\tilde{t}}}
\def\tix{{\tilde{x}}}
\def\tiF{{\tilde{F}}}

\begin{document}

\preprint{}

\title{A Simple Model for Sheared Granular Layers}

\author{Akira Furukawa and Hisao Hayakawa}
\affiliation{%
Department of Physics, Graduateschool of Science, 
Kyoto University, Kyoto 606-8502
}%
\date{\today}

\begin{abstract}
We phenomenologically investigate stick-slip motion 
of sheared granular layers. 
Our phenomenology is constructed in the context of 
nucleation-and-growth of the fluidized area 
which is triggered by collapsing of stress chains. 
Based on this picture, we give a simple friction model 
by introducing the degree of the fluidization.  
It is found that 
the present model can successfully reproduce major features of 
the experimental results reported by Nasuno {\it et al.} 
(Phys. Rev. E {\bf 58}, 2161 (1998)) with quantitatively good agreement.   
\end{abstract}
\pacs{45.70.-n, 45.70.Vn}

\maketitle

Granular friction is one of the central topics of granular science 
\cite{Jaeger-Nagel-Behringer1,Jaeger-Nagel-Behringer2}. 
However, our understanding is still not enough both in microscopic level 
(friction among the individual particles) and in macroscopic level. 

Previously, Nasuno and his co-workers reported 
the detailed experimental studies of stick-slip motion 
observed in sheared granular layers \cite{Nasuno-Kudrolli-Gollub}.  
Their experimental method is very simple and 
is similar to those used in studies of solid-on-solid friction
\cite{Baumberger1,Baumberger2,Baumberger3}. 
That is, the cover plate with the mass $M$ is placed on the granular layer,  
and then it is pushed by a spring (spring constant $k$) 
which is driven at a constant speed $V$.  
Despite this simplicity, 
the obtained results reveal many remarkable things 
which are distinguishable from earlier works on friction phenomena 
in several points.   
Here, lets us summarize main results of their experiment: 
(i) By the fast and sensitive measurements 
of the cover plate on the granular layers,  
they could determine the friction force 
as a function of the instantaneous sliding velocity. 
The resultant friction force is found to be a multivalued function 
of the sliding velocity during the slip event. 
Then characteristic hysteresis is observed. 
(ii) As the driving speed $V$ is increased, 
the stick-slip motion gradually changes into 
the oscillatory motion, where the inertia of the cover plate becomes dominant. 
Furthermore, at a critical driving speed $V_c$ 
the transition to the steady sliding motion occurs. 
For small $k$ the transition exhibits pronounced discontinuity, 
and large fluctuations can be observed 
in the vicinity of the transition point. 
(iii) In the stick-slip regime, 
the observed instantaneous velocity during the slip event 
exhibits almost universal behavior for different driving speed $V$ 
when $k$ and $M$ are fixed.   
(iv) Their measurements of vertical motion of the cover plate indicates that 
the dilatancy might play a crucial role in the granular friction. 
Above all, even within the sticking interval, a small creep occurs 
due to the localized microscopic rearrangement of the granular particles. 
From their detailed observation, they ascribed the observed stick-slip motion  
to repeated ``fluidization'' and ``solidification'' 
of the granular layers. 
(However, the term ``fluidization'' might not be adequate   
because in the experiment the cover plate moves at most 
for a distance of a few particles.)

Motivated by the experiments by Nasuno {\it et al.}
\cite{Nasuno-Kudrolli-Gollub}, several researchers have proposed simple models 
to understand the experimental results
\cite{Hayakawa,Lacombe-Zapperi-Herrmann}. 
Surely, their works succeeded in reproducing some aspects of the experiment, 
but, as a whole, further consideration seems to be needed \cite{comment}. 
In this paper 
we also propose a simple phenomenology for sheared granular layers. 
In order to describe an internal state of the granular layer  
we introduce the degree of the fluidization $\theta$, 
similarly as Refs. \cite{Hayakawa,Carlson-Batista}, 
but in the different context.  
Our model is constructed 
by interpreting the fluidization of the granular layers as  
nucleation-and-growth of the fluidized area 
which is triggered by the collapsing of stress chains.  
As shown in the following,  
the present model can successfully reproduce major features of 
the experimental results reported by Nasuno {\it et al.} 
\cite{Nasuno-Kudrolli-Gollub} with quantitatively good agreement. 
Because of the considerable difficulty from the first principle approach, 
there is still no reliable general theory 
which can explain extensive aspects of granular dynamics.  
So, we believe that the phenomenological approach is useful 
for the present stage of granular physics.

\begin{figure}[h]
\includegraphics[width=.8\linewidth]{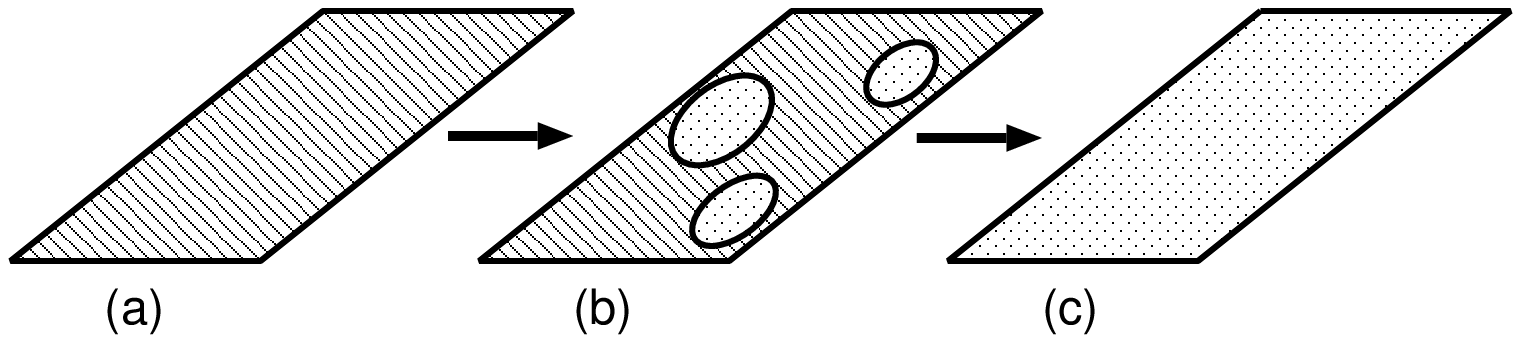}
\caption{
Schematic illustration of our phenomenology 
seen from the top of the granular layer. 
(a) At the maximum friction $F_{\rm max}$. 
(b) When the applied force exceeds $F_{\rm max}$, 
the granular layer begins to fluidize. 
Our phenomenology assumes this fluidization proceeds as nucleation-and-growth, 
and the nuclei is attributed to the points 
at which the break-down of stress chains initially take place. 
The fluidized areas represented by dots dynamically coexist 
with the non-fluidized areas represented by diagonal lines. 
(c) The whole region becomes fluidized.  
}
\label{schematic-diagram}
\end{figure}

We now introduce a simple friction model 
for the sheared granular layers. 
Here, using a schematic illustration shown in Fig. \ref{schematic-diagram}, 
we shall explain our phenomenology as follows. 
(i) During the sticking period, 
the friction force is balanced with the applied force 
from the pulled spring.
In the meantime, the stress chains develop in the granular layer 
so that they sustain the increased applied force, 
which should be a microscopic origin of increasing of the friction force. 
Eventually, the system reaches to 
the state exhibiting the maximum static friction $F_{\rm max}$ 
(Fig. \ref{schematic-diagram}(a)), 
where the granular layer can no longer sustain more external force.  
(ii) When the applied force exceeds the maximum static friction 
$F_{\rm max}$, the cover plate begins to slide. 
Simultaneously, the collapsing of the stress chains should trigger 
the fluidization of the granular layers,  
giving rise to pronounced decrease of the friction force.    
It is known that the network of the stress chains spreads heterogeneously 
in the granular layer, resulting in the heterogeneous stress transmission. 
Instead of imaging that the fluidization does not happen uniformly, 
hence, we now ascribe it to nucleation-and-growth of the fluidized area 
(Fig. \ref{schematic-diagram}(b)).  
In this paper ``fluidized area'' is interpreted as a region 
in which the sustained stress is smaller than 
that in the region with fully developed stress chains.  
This is due to the break-down of the stress chains. 
Since the fluidization is a local process at the fluidization front,
it is assumed that the fluidization fronts propagate with a constant velocity
independently of the cover plate.
(iii) In a short time, 
the whole area becomes fluidized (Fig. \ref{schematic-diagram}(c)), 
and then the friction force exhibits almost saturated value. 
(iv) Finally, the cover plate ceases to move, 
and then the stick regime recovers again.  
In the experiment \cite{Nasuno-Kudrolli-Gollub}, 
this re-sticking process suddenly occurs 
with pronounced decrease of the friction force.

We are at the position to construct our friction model 
based on the above mentioned picture. 
To begin with, we introduce the degree of the fluidization $\theta$  
with the aid of the simple picture of phase change,  
instead of considering the microscopic dynamics of the granular particles.   
Let us imagine, for an example, 
a magnetic system in an external magnetic field, 
where constituent spins point to the preferable direction.   
If the external field is turned over, 
the whole region of the system becomes unstable, 
and then stable phase, with spins pointed to the opposite direction, 
nucleates and grows. 
This nucleation-and-growth for non-conserved order parameter system 
has been known as phase change (or phase switching) 
and is well described by Avrami-Kolmogolov theory  
\cite{Avrami,Kolmogolov,Ishibashi-Takagi}. 
The velocity of propagating front of new phase is a constant, 
so that in the early stage of growth process,  
the volume fraction of the new phase grows as $t^{d+1}$, 
with $d$ being a spatial dimension, 
in the case of constant nucleation rate.  
On the other hand, the volume fraction grows as $t^d$ 
in the case where there are only latent nuclei 
at the beginning but no new nucleation during the growth regime.       
In this paper, using this analogous situation, 
the fluidization is interpreted as a
2-dimensional phase change with latent nuclei at the beginning. 
Thus, $\theta(t)$ is introduced as the volume fraction of the fluidized area, 
and is assumed to be given by the following simple form 
\EQ
\theta(t)=1-\exp(-A(t)), \label{degree}
\EN
where $A(t)$ is the so-called Avrami's extended volume fraction  
\cite{Avrami,Kolmogolov,Ishibashi-Takagi}, which is given by 
$A(t)=(\alpha \tau)^{2}$ in the present case, 
where $\alpha^{-1}$ is some characteristic time 
relating to the microscopic mechanism of the fluidization.  
The saturated state is realized (Fig. \ref{schematic-diagram}(c)) 
when $\alpha \tau\gg 1$. 
Here $\tau=t-nT$, where $n$ is a number, and $T$ 
is the period of stick-slip event, respectively. 
We set $t=0$ just at the beginning of the first slip event. 
The exponent 2 may be replaced by some fractal number 
reflecting the complicated structure of the growing front.

An equation of motion of the cover plate is given by 
\EQ
M\ddot{x}=k\delta x-F,  \label{eq-motion}
\EN 
where $\delta x=Vt-x$ 
is the displacement from the natural length of the spring.  
When the fluidization proceeds as in Fig. \ref{schematic-diagram}(b), 
the friction becomes smaller for the part of the fluidized area.  
Therefore, 
we propose the following dynamic friction force $F_d$ in terms of $\theta$
\EQ
F_{d}=F_0+F_1[1-\theta(t)]+b\dot{x},  \label{d-friction}
\EN
where $F_1[1-\theta(t)]$ represents 
the decreasing part of the friction owing to collapsing of the stress chains, 
and $F_0+F_1=F_{\rm max}$.  
Practically, in an usual solid-on-solid friction, 
it is well known that the friction force is 
proportional to the real contact area rather 
than to the apparent contact area \cite{Persson}. 
Eq. (\ref{d-friction}) can be understood as its modified version. 
Here we add the velocity dependent friction force $b\dot{x}$,  
which is needed to describe correct transition behavior.  
It must be mentioned that the present model 
as well as the models in Refs. \cite{Hayakawa,Lacombe-Zapperi-Herrmann} 
cannot describe the re-sticking process, 
and therefore Eq. (\ref{d-friction}) is valid only for ${\dot x}>0$. 
So, in the sticking period 
we must use the static friction as $F_{s}=k{\delta x}$.

We make Eq. (\ref{eq-motion}) a dimensionless form 
by using the variables,  
$\tit=\sqrt{{k}/{M}}t$, $\tix=({k}/{Mg})x$, and $\tiF={F}/{Mg}$.  
Then, the equation of motion (\ref{eq-motion}) becomes 
\EQ
{\ddot \tix}=a\tit-\tix-\tiF, \label{scaled-eq-motion}
\EN
where $a={V}/{g}\sqrt{{k}/{M}}$, 
and the friction force is written as  
\EQ
\tiF=\left\{
\begin{array}{ll}
\tiF_d=\tiF_0+\tiF_1[1-\theta(\tit)]+\frac{b}{\sqrt{Mk}} {\dot \tix}
~~~~({\dot \tix}>0),\\
\tiF_s=a\tit-\tix~~~~(a\tit-\tix<\tiF_{\rm max}).
\end{array}
%\tiF_d&=&\tiF_0+\tiF_1[1-\theta(\tit)]+\frac{b}{\sqrt{Mk}} {\dot \tix}
%~~~~({\dot \tix}>0)\label{dynamic}\\
%\tiF_s&=&a\tit-\tix~~~~(a\tit-\tix<\tiF_{\rm max}) \label{static}
\right.\label{friction}
\EN
In the following numerical analysis, we set $b=0.392 ({\rm kg/s})$, 
$\tiF_0=0.44$, and $\tiF_1=0.22$, 
which are estimated from the experimental results. 
${\tilde\alpha}=\alpha\sqrt{M/k}$ is an adjustable parameter 
and we set ${\alpha}=146.99$ (${\rm 1/s}$).

\begin{figure}[htb]
\includegraphics[width=0.8\linewidth]{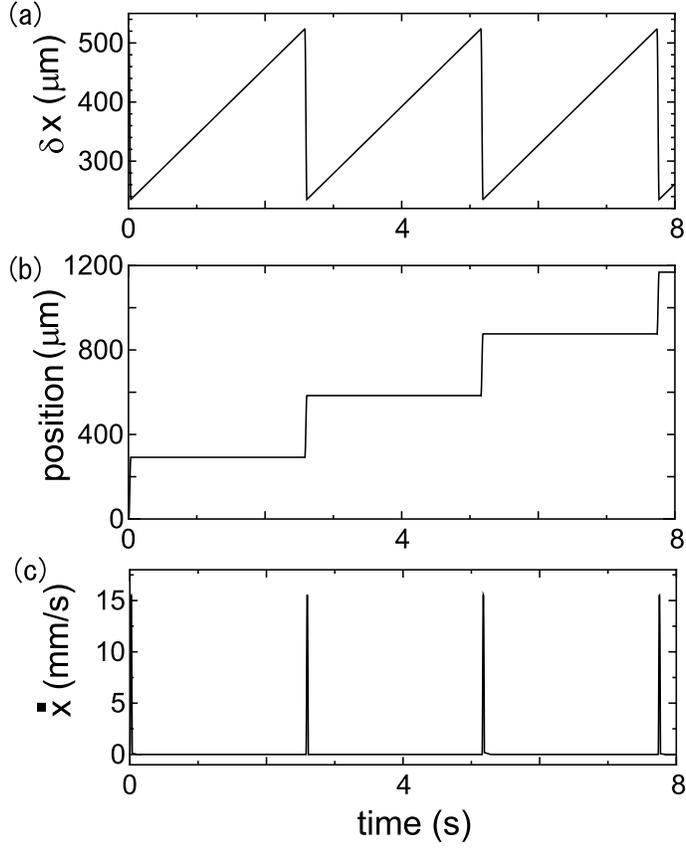}
\caption{
The numerically calculated 
(a) deflection $\delta x(t)$, 
(b) position $x(t)$, and 
(c) instantaneous velocity ${\dot x}(t)$, 
for $k=134.7$(N/m), $M=10.9$(g), and $V$=113.33($\mu$m/s). 
Here the parameters used are 
the same as those in Fig. 3 of Ref. \cite{Nasuno-Kudrolli-Gollub}. 
}
\label{stick-slip}
\end{figure}

\begin{figure}[h]
\includegraphics[width=0.75\linewidth]{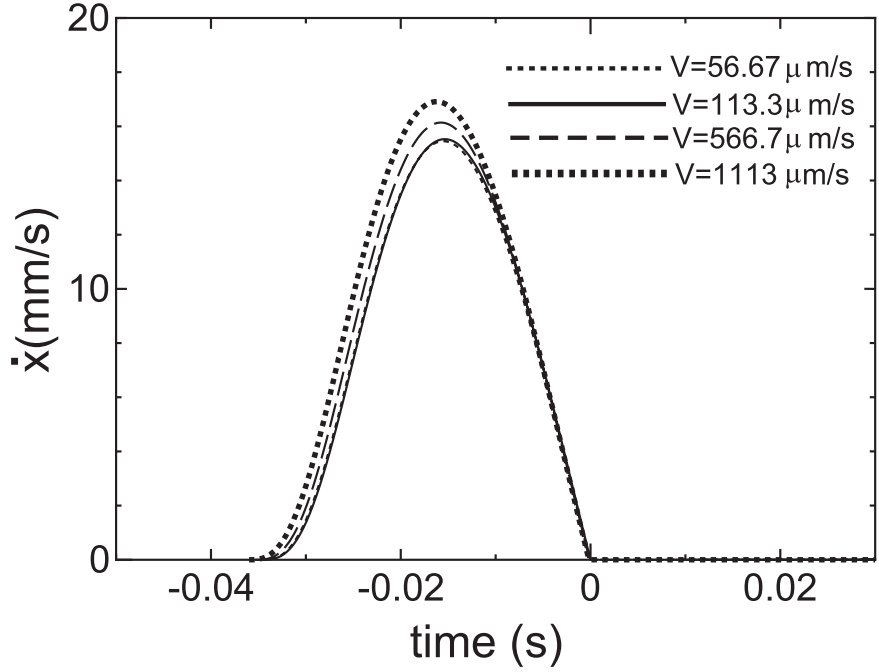}
\caption{
Instantaneous velocity of the cover plate $\dot x$ for various value of $V$ 
at $k=135$(N/m) and $M=10.90$(g). The parameters used are 
the same as those in Fig. 11 of Ref. \cite{Nasuno-Kudrolli-Gollub}. 
}
\label{velocity}
\end{figure}

\begin{figure}[h]
\includegraphics[width=0.85\linewidth]{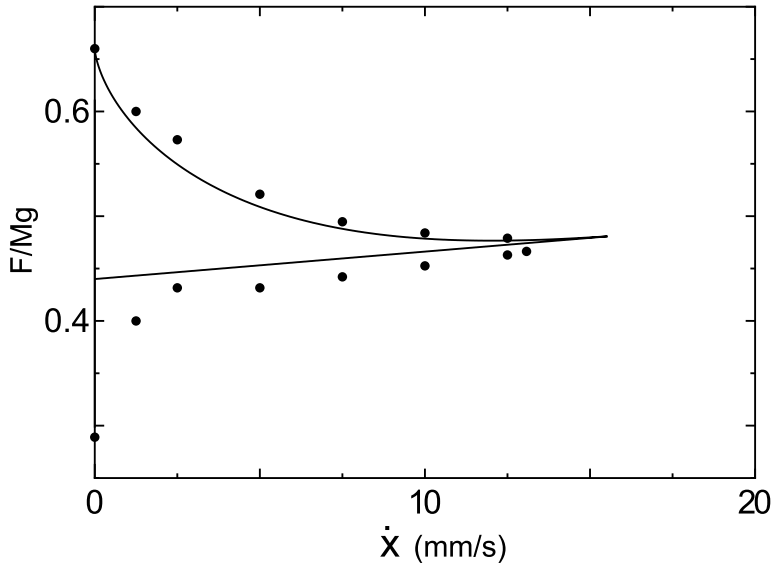}
\caption{
The normalized friction force $F/Mg$  
as a function of instantaneous velocity $\dot x$ 
for $k=134.7$(N/m), $M=10.9$(g), and $V=113.33(\mu$m/s). 
The filled circles represent the experimental results, 
which are taken from Fig. 13 of Ref. \cite{Nasuno-Kudrolli-Gollub}. 
}
\label{hysteresis loop}
\end{figure}

Eq. (\ref{scaled-eq-motion}) with Eq. (\ref{friction}) 
can be numerically integrated in a light manner.  
The obtained typical stick-slip motion is shown in Fig. \ref{stick-slip},
where the numerically calculated (a) deflection $\delta x(t)$, 
(b) position $x(t)$, and (c) instantaneous velocity ${\dot x}(t)$ 
are presented, respectively.  
In the following, we shall consider our numerical results in some detail.  
Fig. \ref{velocity} represents instantaneous velocity 
for various $V$ at $k=134.7$(N/m) and $M=10.90$(g), 
which corresponds to Fig. 11 in  Ref. \cite{Nasuno-Kudrolli-Gollub}. 
Similarly to the experiment \cite{Nasuno-Kudrolli-Gollub},  
we find that the instantaneous velocity is hardly changed against $V$. 
This is entirely due to our choice of the form of $F$.  
That is, 
our friction model is intrinsically insensitive to the driving speed $V$.
Now, it must be recalled that we have constructed our phenomenology 
imaging that the fluidization occurs as nucleation-and-growth.  
Then, the degree of the fluidization $\theta$ has been introduced 
as Eq. (\ref{degree}).   
There, we have implicitly assumed that the motion of the cover plate  
is important only as a trigger of the break-down of the stress chain,  
and then the resultant collapsing dynamics of the stress chains proceeds  
almost independently of the motion of the cover plate.   
The experimental result might support our assumption indirectly. 
From Eq. (\ref{friction}),    
we can calculate the friction force as a function of ${\dot x}$. 
The result is shown in Fig \ref{hysteresis loop}, 
where the filled circles represent experimental results 
\cite{Nasuno-Kudrolli-Gollub}.   
In the early stage of the slipping event ($Vk/F_1\alpha^2\ll t\ll 1/\alpha$),  
the velocity $\dot x$ is well approximated by 
\EQ
{\dot x}\cong \frac{F_1\alpha^2}{3M}t^3.   
\EN
Hence, the friction force decreases as 
\EQ
F\cong F_{\rm max}+b{\dot x}-(3M\sqrt{F_1}\alpha)^{\frac{2}{3}}{\dot x}^{\frac{2}{3}}. 
\EN 
Though the above expressions depend on the detailed structure 
of the fluidization, Eq.(\ref{degree}),  
these behavior in the early stage of the slip event is also well consistent 
to the experimental results \cite{Nasuno-Kudrolli-Gollub}.

\begin{figure}[h]
\includegraphics[width=0.7\linewidth]{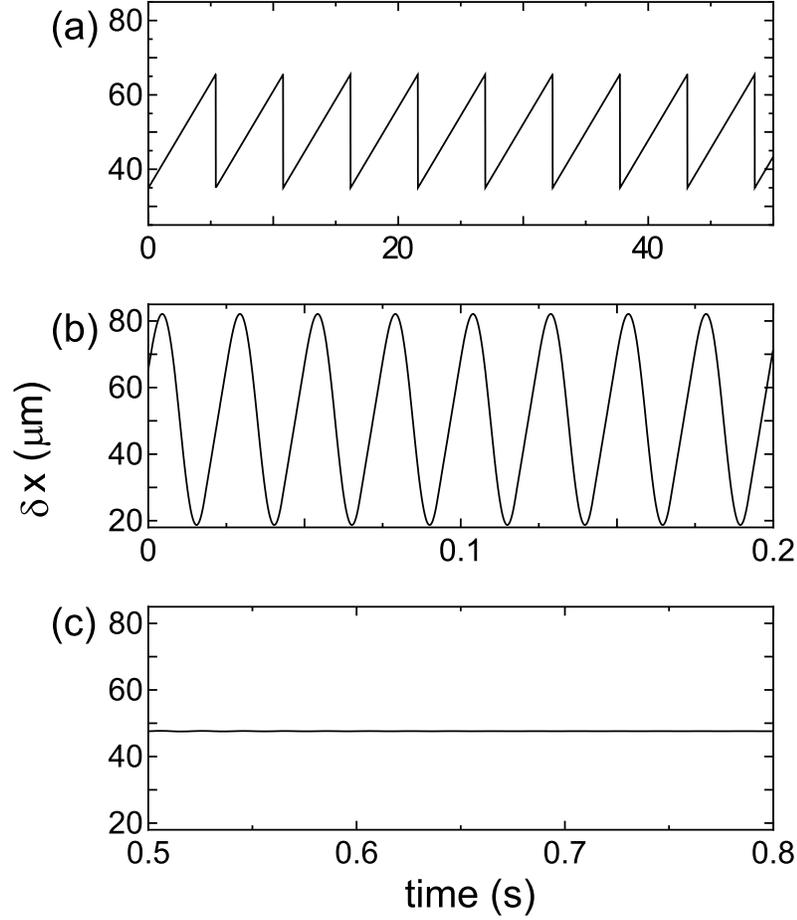}
\caption{Deflection $\delta x(t)$ for various $V$ at $k=1077$(N/m) and $M=$10.9(g). (a) Typical stick-slip motion is observed at $V=5.67(\mu$m/s). 
(b) Oscillatory stick-slip behavior at $V=5.67$(mm/s). 
(c) Steady-sliding motion at $V=12.4$(mm/s), 
where we do not show the transient damped oscillatory motion observed for 
$t< 0.5$(s). The parameters used here is the same as those 
in Fig. 7 of Ref. \cite{Nasuno-Kudrolli-Gollub}, 
except for (c), where we use slightly different value of $V$. 
}
\label{transition1}
\end{figure}

\begin{figure}[h]
\includegraphics[width=0.7\linewidth]{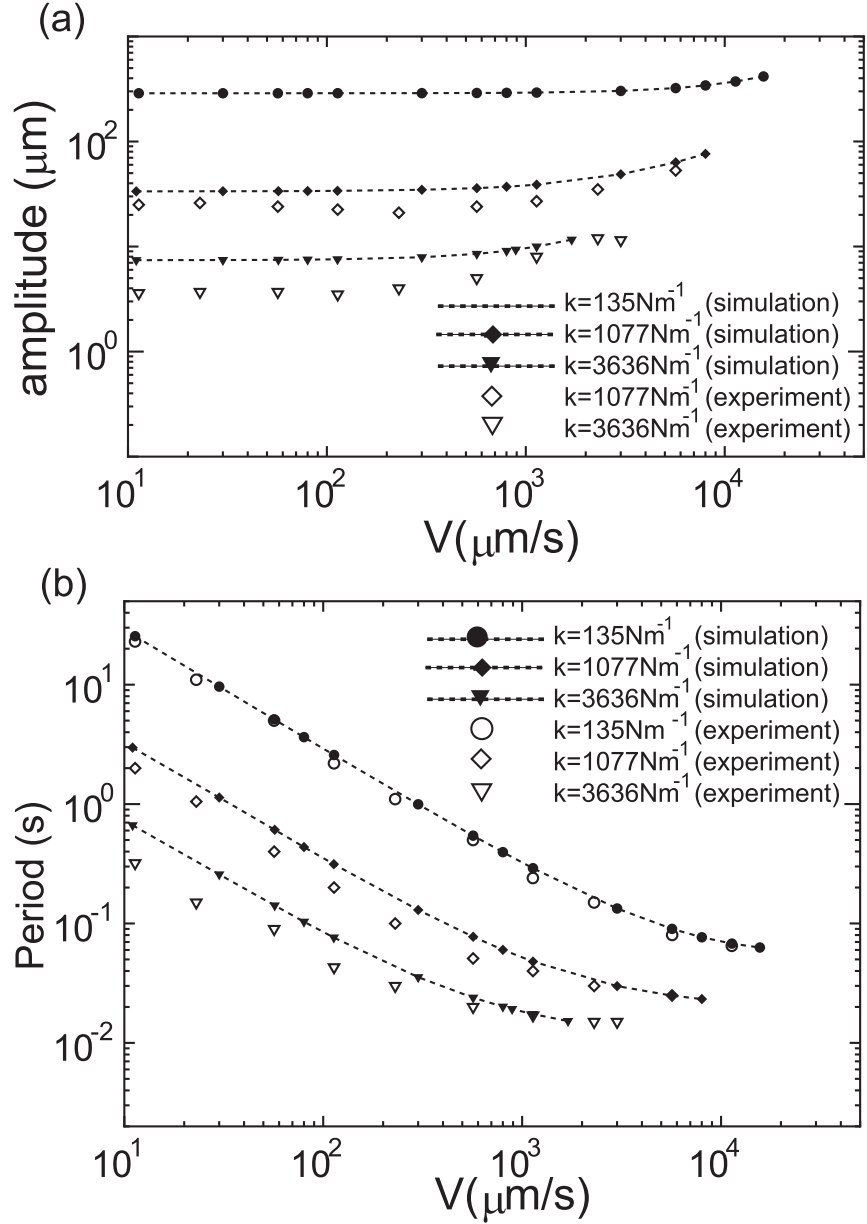}
\caption{
(a) Amplitude of the deflection $\delta x$ 
of stick-slip motion as a function of $V$ for various $k$.  
(b) Period of stick-slip motion as a function of $V$ for various $k$.  
The open marks are taken from Fig. 8 
in Ref. \cite{Nasuno-Kudrolli-Gollub}, 
where the data for $k=135$(N/m) is not available for $\delta x$.
}
\label{transition2}
\end{figure}

Next, we present numerical results for the dynamic transition behavior 
observed in the experiment \cite{Nasuno-Kudrolli-Gollub}.  
As increasing the driving speed $V$,  
the inertia of the cover plate becomes crucial.  
As a result, the stick-slip motion becomes smoother, 
and then the oscillatory motion is observed (Fig. \ref{transition1}(b)).  
Moreover, above a certain critical value $V_c$,  
the oscillatory motion changes into the steady sliding motion 
(Fig. \ref{transition1}(c)).  
The period and the amplitude of stick-slip motion 
against the driving velocity $V$ are shown in Fig \ref{transition2}, 
in which the dotted line ceases at the point 
where the stick-slip behavior vanishes.   
In an usual stick-slip regime, the period behaves as $V^{-1}$ 
and the amplitude of $\delta x(t)$ is almost constant. 
But, as $V$ increases,  
the modulation from such behavior becomes pronounced 
due to the dominance of the inertia of the cover plate. 
Comparing with the experimental result by Nasuno {\it et al.},   
we find that our obtained transition behavior shows 
good agreement with quantitative sufficiency.

In this paper we have phenomenologically investigated stick-slip friction of 
sheared granular layers reported by Nasuno {\it et al.} 
\cite{Nasuno-Kudrolli-Gollub}. 
Our model constructed 
in the context of nucleation-and-growth of the fluidization can reproduce 
major aspects of the experimental results. 
We consider that the present idea might be also useful 
to the investigation of the dynamics of earthquake. 
However, our model is still insufficient to the 
complete understanding of the experimental works.  
(i) We can say nothing about the re-sticking process.  
In the experiment \cite{Nasuno-Kudrolli-Gollub}, 
the remarkable decrease in the friction force is observed 
just before re-sticking. 
At this time, the dilation of the granular layers ends. 
This observation indicates that 
the vertical motion of the granular layers might play a crucial role 
to the re-sticking. 
(ii) Our present approach is mean-field like, 
so that the fluctuation effect is not included.  
In the experiment \cite{Nasuno-Kudrolli-Gollub}, 
such an effect surely leads to  
intriguing chaotic stick-slip motion for very low $V$ and large $k$.   
These problems are left for future works\cite{future-work}.

One of the present author  
A.F. thanks to Prof. Takao Ohta and Prof. Yoshitsugu Oono 
for valuable comments. 
The major parts of the present paper had been done in 1998,  
when A.F. was an undergraduate student of 
Department of Physics at Kyoto university.  
He also thanks to Dr. Tsuyoshi Mizuguchi and Dr. Ryoichi Yamamoto 
for their useful discussions and encouraging comments of those days. 
This study is partially supported by Grant-in-aid for JSPS Fellows 
and by Grant-in-Aid for Scientific Reserch (Grant No. 15540393) 
of Ministry of Education Science of Culture, Japan. 
 
\newpage

\end{document}